\documentclass{elsart}
\usepackage{graphicx,harvard,amssymb}
\journal{New Astronomy}
\input psfig.sty

\def\astrobj#1{#1}

\begin{document}
\begin{frontmatter}

\title{The Spectroscopic Orbits of Three Double-lined Eclipsing Binaries: I. \astrobj{BG~Ind}, \astrobj{IM~Mon}, \astrobj{RS~Sgr}}

\author{V. Bak{\i}\c{s}$^1$\corauthref{cor1}},
\author{H. Bak{\i}\c{s}$^1$},
\author{S. Bilir$^2$},
\author{F. Soydugan$^1$},
\author{E. Soydugan$^1$},
\author{O. Demircan$^1$},
\author{Z. Eker$^{3,4}$},
\author{E. Yaz$^2$},
\author{M. T\"{u}ys\"{u}z$^1$},
\author{T. \c{S}eny\"{u}z$^1$}

\address{$^1$\c{C}anakkale Onsekiz Mart University Physics Department and Ulup{\i}nar Observatory, Terzio\u{g}lu Kamp\"{u}s\"{u}, 17020, \c{C}anakkale, Turkey \\
$^2$Istanbul University Science Faculty, Department of Astronomy and Space Sciences, 34119, University-Istanbul, Turkey \\
$^3$Akdeniz University, Faculty of Scince and Art, Physics Department 07058 Campus, Antalya, Turkey\\
$^4$T\"UB\.ITAK National Observatory, Akdeniz University Campus, 07058 Antalya, Turkey}

\corauth[cor1]{Corresponding author. Tel.: +90 286 2180019; Fax: +90
268 2180533;\\ \textit{Email address}: bakisv@comu.edu.tr}

\maketitle

\label{firstpage}

\begin{abstract}
We present the spectroscopic orbit solutions of three double-lines eclipsing binaries, BG Ind, IM Mon and RS Sgr. The first precise radial velocities (RVs) of the components were determined using high resolution echelle spectra obtained at Mt. John University Observatory in New Zealand. The RVs of the components of BG Ind and RS Sgr were measured using Gaussian fittings to the selected spectral lines, whereas two-dimensional cross-correlation technique was preferred to determine the RVs of IM Mon since it has relatively short orbital period among the other targets and so blending of the lines is more effective. For all systems, the Keplerian orbital solution was used during the analysis and also circular orbit was adopted because the eccentricities for all targets were found to be negligible. The first precise orbit analysis of these systems gives the mass ratios of the systems as 0.894, 0.606 and 0.325, respectively for BG Ind, IM Mon and RS Sgr. Comparison of the mass ratio values, orbital sizes and minimum masses of the components of the systems indicates that all systems should have different physical, dynamical and probable evolutionary status.
\end{abstract}


\end{frontmatter}

\section{Introduction}

The theoretical knowledge and the understanding of fine structure and evolution of single stars could only be improved if their physical parameters are obtained with a sufficient precision. Accessing reliable accurate physical parameters, however, requires careful study of eclipsing binary stars using spectroscopy and photometry. In this paper, we have observed and studied three Algol-type eclipsing binaries (BG~Ind, IM~Mon and RS~Sgr) of which apparently have not yet studied with a sufficient accuracy and thus their physical parameters in the literature are absent nor accurate to understand their orbits and physical natures. Being one of a series (Bak{\i}\c{s} et al., 2007; Bak{\i}\c{s} et al., 2008a; Bak{\i}\c{s} et al., 2008b)  of publications, here now we announce the optical spectra and spectroscopic elements of BG~Ind, IM~Mon and RS~Sgr.

\subsection{BG~Ind}

BG~Ind (also HD 208496 = HR 8369 = SAO 247247) is listed as a possible eclipsing binary in the fourth edition of the Bright Star Catalogue (Hoffleit \& Jaschek 1991), while Manfroid \& Mathys (1984) confirmed the eclipsing nature of BG~Ind and established an orbital period of 1.464047 days from its photometric observations. The first spectral analysis of the system by Andersen, Jensen \& Nordstr\"om (1984) showed that HD 208496 is a double-lined eclipsing binary with F3V spectral type. Kholopov et al. (1987) designated it as BG~Ind in the 68th Name-list of Variable Stars and classified it as EA (Eclipsing Algol).

Although it is a bright system, due to its southern declination, there are only a few studies discussing the nature of the system. van Hamme
\& Manfroid (1988) analyzed the system with ground-based photometric light curves in Str\"{o}mgren $uvby$ bands. Using the two radial velocities (RVs) given earlier by Andersen et al. (1984), van Hamme \& Manfroid (1988) were able to estimate the mass and radius of the component stars as 1.41$M_{\odot}$ and 2.22$R_{\odot}$, 1.20$M_{\odot}$ and 1.60$R_{\odot}$, for the primary and secondary, respectively. BG~Ind was also observed by the Hipparcos satellite (Perryman et al. 1997). The Hipparcos light curve shows that the magnitude of the system at maximum brightness is $V_{max}=6.11$ mag and the depth of the primary eclipse is 0.25 mag.

\subsection{IM~Mon}

The variability of IM~Mon (also HD 44701 = HIP 30351 = SAO 133189) was discovered by Gum (1951), who published the first photometric light curve. The light curve of the system obtained by Gum (1951) is asymmetric with scatter and implies a system consisting of two detached components sufficiently close to each other to exhibit proximity effects. Gum (1951) analyzed the light curves without spectroscopic information and gave the relative dimensions of the component stars. Sanyal \& Sinvhal (1964) presented new photometric observations of IM~Mon. Their light curves in B and V filters show the same asymmetry as seen in the light curve of Gum (1951). According to Sanyal \& Sinvhal (1964), the scatter seen in the light curve of IM~Mon is not random but systematic due to intrinsic variability of the cooler component.

The first spectroscopic observations of the system were made by Pearce (1951). Analyzing 18 spectra, Pearce (1951) gave the spectroscopic orbital elements for the first time and the absolute dimensions of the components. According to the analysis of Pearce (1951), IM~Mon consists of two main sequence stars with spectral types of B4 and B7 and masses of 6.8M$_\odot$ and 4.5M$_\odot$ for the primary and secondary, respectively. Cester et al. (1978) gave the updated light curve elements using the photometric data of Gum and Sanyal \& Sinvhal (1964). The most recent V-band photometric observations of IM~Mon were made by Shobbrook (2004), which were not attemped to be solved. Our present study attempts to obtain precise orbital elements of the system to study the components in more detail, especially the secondary components which was so called responsible for the scatter in the light curve.

\subsection{RS~Sgr}

RS~Sgr (also HD 167647 = HIP 89637 = SAO 209959) is bright southern eclipsing binary. Roberts (1901) confirmed the variability of the system and determined the first times of minimum light. In the recent orbital period study of RS~Sgr by Cerruti \& de Laurenti (1990), the orbital period of the system is given as 2.41568311 days. The spectroscopic studies of the system were started in the beginning of 20th century. The latest spectroscopic study related to RS~Sgr was made by Ferrer \& Sahade (1986). Analyzing 46 spectra of the system, Ferrer \& Sahade (1986) gave the spectroscopic orbital elements and masses of the components using the orbital inclination given by Baglow (1948). The latest study of the absolute dimensions of the components were given by Cerruti \& de Laurenti (1990). According to the analysis of Cerruti \& de Laurenti (1990), RS~Sgr consists of two main sequence stars with spectral types of B5 and A2, masses of 7.18M$_\odot$ and 2.41M$_\odot$, radius of 5.11R$_\odot$ and 4.10R$_\odot$, for the primary and secondary, respectively. Here, we re-study the system spectroscopically.

\section{Observations}

The spectra of BG~Ind, IM~Mon and RS~Sgr were taken with High Efficiency and Resolution Canterbury University Large Echelle Spectrograph (HERCULES) of the Department of Physics and Astronomy, New Zealand. It is a fibre-fed \'{e}chelle spectrograph, attached to the 1-m McLellan telescope at the Mt John University Observatory (MJUO), Lake Tekapo ($\sim$ 43$^{\circ}$ 59$^{m}$ S, 174$^{\circ}$ 27$^{m}$ E). The spectrograph is mounted on a stable optical bench and enclosed within a vacuum vessel located in a thermally insulated room. Light from the Cassegrain focus can be fed to the spectrograph using a 20-m optical fibre of selectable cross-section (Hearnshaw et al. 2002). The spectrograph is designed to provide continuous wavelength coverage 380 to 880 nm in over 80 \'{e}chelle orders. Two resolving powers of R=41000 and 70000 are possible. Choices of resolving powers are made possible by selecting one of three optical fibres with different sizes. Fibre 1, which is a 100-$\mu$m fibre without slit giving a moderately high resolving power of 41000, was selected for the observations of this project. The angular size of the fibre core projected on to the sky (4.5 arcsec) is not less than the median seeing at MJUO which was reported to be about 3.5 arcsec (Hearnshaw et al. 2002). Fibre 2 and 3 have a resolving power of R=70 000 however they require better seeing conditions ($\theta$$\sim$2 arcsec) to obtain spectra with good signal-to-noise (S/N). A more detailed description of the spectrograph can be found in Hearnshaw et al. (2002) and Skuljan, Ramm \& Hearnshaw (2004).

All spectra have been collected using the 4k$\times$4k Spectral Instruments 600S CCD camera. The pixel size of the camera is 15-$\mu$m and covers the complete spectral range. All the spectra presented in this paper are obtained in the same observing session (August\,2006). The total number of spectra for BG~Ind, IM~Mon and RS~Sgr are 41, 23 and 35, respectively. The journal of observations including observing times, mean S/N ratios at 550 nm and exposure times are given in Tables~1, 2 and 3.

For all observations, comparison spectra of thorium-argon arc lamp for wavelength calibration are recorded before and after each exposure. The dispersion solution was interpolated between the two arc lamp spectra for the exact time of the stellar observation which is flux weighted mean obtained from the exposure meter readings during the observation. Exposure time of each stellar spectrum was determined to achieve desired S/N ratio which were about 100 around 550 nm for each star. A set of white lamp images was taken every night for flat fielding.

All spectra were reduced with the HERCULES Reduction Software Package (HRSP) developed in the Department of Physics and Astronomy, University of Canterbury. The procedure of the reduction is a standard for \'{e}chelle spectra (Skuljan 2004).

The normalization procedure of our reduced spectra are as follows: we first determined the shape of the stellar continuum at the order
being studied by fitting second or third order polynomials to the stellar continuum depending on the shape of the continuum and then
divided by the fitted polynomials.

\section{Radial Velocities and Spectroscopic Orbits}

\subsection{BG~Ind}

Relatively sharp lines of the components of BG~Ind allow measurements of Doppler shifts of the spectral lines by simple Gaussian fittings. However, a selection of spectral order plays an important role for a reliable RV measurement since the spectral features of the components must be far enough away from each other to avoid blending. For this reason, among the 89 spectral orders available in each of our \'{e}chelle spectrum we selected the wavelength region 6050--6150 \AA\,(the absolute order number: 93) where the Ca I (6102.723\AA) lines of the components are strong enough for reliable RV measurement (see Fig.~1). The Gaussian fitting to the lines was performed using the splot task of IRAF. In case of blending of lines at near conjunction times (see Table~1), deblend function was used to fit two Gaussians to the blended lines. However, at times when the spectral feature of one of the component star is behind the other or very strongly blended at the conjunction time, the RV of the former was not measured. The laboratory wavelength of Ca I line used for the determination of the V$_\gamma$ velocity are taken from the NIST database (Ralchenko et al., 2008). The heliocentric correction are applied to all RVs measured before using them in the orbital solution. The final RVs are given in Table~1.

van Hamme \& Manfroid (1988) have determined the orbital period of BG~Ind as 1.464069 days from a set of light curves obtained at different epochs. This orbital period was assumed to be reliable and used during our analysis of orbital solution.

With the RVs obtained from the Gaussian fitting, pure Keplerian orbital solution was obtained by least-squares fitting. The orbital period was kept fixed, while the eccentricity $e$, longitude of periastron $\omega$, velocity semi-amplitudes $K_{1,2}$, systemic velocity $V_{\gamma}$ and conjunction time have been converged in the least-squares solution. After a few attempts, the eccentricity
is found to be very small ($e$=0.002) compared to its uncertainty (0.007). Therefore, the final feasible spectroscopic orbit was
adopted assuming a circular orbit. The adopted orbital parameters are given with uncertainties in Table~4. The theoretical Keplerian
orbital fitting to the O-C residuals is shown in Fig.~4.

\subsection{IM~Mon}

Our spectra of IM~Mon confirm its early spectral type (B5V) reported in SIMBAD database. Apart from the Balmer lines, the most prominent
lines in the spectra are He I lines and the Mg II line at 4481 \AA. Among the He I lines, 6578 \AA\,and 5875 \AA\,are relatively stronger than the rest. The region where the He I 5875 \AA\,line is located includes telluric lines of water vapor (H$_2$O) which strongly blends with the stellar line. Therefore, the He I 6675 \AA\,line have been chosen for reliable RV measurements. The spectral region of He I 6678 \AA\,line is given in Fig.~2 with orbital phase indicated for each spectrum.

From the component spectral lines in Fig.~2, it can be seen that the line profiles of the components are broadened primarily by the rotation. Thus, the blending of the lines for this close binary are relatively frequent during the orbital motion. We, therefore, used
two-dimensional cross-correlation technique (the code {\sc TODCOR}). The algorithm of {\sc TODCOR} was developed by Zucker \& Mazeh
(1994) and has been efficiently applied to multiple-component spectra of late-type double-lined spectroscopic binaries (Latham et
al. 1996; Metcalfe et al. 1996, among others) and more recently to short-period early-type binaries (Gonz\'{a}les \& Lapasset 2003;
Southworth \& Clausen 2007). Basically, {\sc TODCOR} calculates two-dimensional cross-correlation function (CCF) from one observed
and two template spectra in re-binned log $\lambda$ space and then locates the maximum of the CCF. We re-binned our observed and template spectra so that each pixel corresponded to 1.6 kms$^{-1}$. The template spectrum of each component was synthetically produced using the appropriate model atmosphere grids of Kurucz (1993). The atmosphere parameters (log$g$, T$_{eff}$) of the components were adopted assuming the primary is B4V and the secondary is B7V spectral type as estimated by Pearce (1951). The projected rotational velocity of the components were determined by comparison of the synthetically broadened spectral lines with the observed spectra. The RVs of the components were computed from the 2-dimensional CCFs with the highest score which is formed by combining the one dimensional CCFs of the shifted and rescaled template spectra with the observed one. The RVs measured are presented in Table~2.

Using the RVs determined by {\sc TODCOR} routine, the orbital solution was performed by least-squares orbital fitting. The orbital period of $P$$=$1.1902424 days, which was taken from the web page ``Up-to-date Ephemerides of Eclipsing Variables'' provided by Kreiner (2004), was fixed while the velocity semi-amplitudes $K_{1,2}$, systemic velocity $V_{\gamma}$ and the conjunction time were converged in the least-squares solution. As in the analysis of BG~Ind, the Keplerian orbital solution of IM~Mon was first made by converging the eccentricity ($e$) and the longitude of periastron ($\omega$) parameters. Once the eccentricity is found to be negligible, we adopted circular orbit in following orbital fittings. The radial velocities of the secondary component at phases $\phi$$=$0.07, 0.56 and 0.44 could not be measured due to strong blending. We also did not use the radial velocities of the components at phases near conjunction ($\phi$$=$0.91, 0.07 for primary, $\phi$$=$0.41 for secondary star) to avoid the Rossiter-McLaughlin effect (Rossiter, 1924) seen at the eclipsed star's line profiles due to rotation. The adopted orbital parameters from the least-square fitting are listed in Table~4 and the orbital fitting to RVs is shown in Fig.~5.

\subsection{RS~Sgr}

Along the optical spectrum of RS~Sgr, relatively strong neutral He lines of the primary component indicate the early spectral type status of the component. Weaker light contribution of the secondary component as well as its later spectral type (Ferrer \& Sahade, 1986) avoid detection of the lines of the cooler component. We found that Mg II (4481 \AA) lines of both component are the most suitable lines for reliable RV measurements.

Applying Gaussian fittings to the individual lines and the use of deblend function of IRAF when the individual lines are blended, as it was done in the case of BG~Ind, allowed us to measure RVs of the components which are given in Table~3.

The least-square orbital solution has been applied to the RVs in Table~3. During the orbital fitting, the orbital period of the systems given by Kreiner (2004) while other orbital elements such as conjunction time T$_{0}$, eccentricity $e$, longitude of periastron $w$, velocity semi-amplitudes K$_{1,2}$ and the systemic velocity V$_\gamma$ were converged. In every iteration the eccentricity of the orbit was found to be zero. Therefore, we assumed the circular orbit in our final adopted solution. The spectroscopic orbital elements of RS~Sgr found in our analysis are given in Table~4 together with their uncertainties.

\section{Final Results}

We presented the orbital elements of three double-lined eclipsing binaries using high resolution spectroscopic data. From the orbital
solutions of the systems investigated in this paper, the following results were obtained:

1- According to the orbital elements of BG~Ind, the mass ratio of the system is $q=$0.894$\pm$0.012 which is 5 per cent higher than the mass ratio given by van Hamme and Manfroid (1988) ($q=$0.85). Moreover, 18 kms$^{-1}$ difference between our V$_\gamma$ velocity and the value given by Andersen et al. (1984) is noteworthy.

2- The spectroscopic solution of IM~Mon imply a discrepancy of 9 per cent between our mass ratio and of Pearce (1951) ($q=$0.665$\pm$0.024). Comparing to ours, ($q=$0.606$\pm$0.023), the difference is larger than the error limits. The discrepancy may
simply be caused by the low resolution spectra used in his study which may cause unreliable RV measurements especially for the secondary component with weak spectral lines.

3- The low mass ratio (0.325$\pm$0.004) of RS~Sgr found from the reliable RV measurements of the components supports a semi-detached binary configuration as suggested by Cerruti \& Laurenti (1990). It should be checked by the analysis of RVs together with light curves.

The physical parameters and the evolutionary status of components of the close binaries studied in present paper will be investigated
using spectral disentangling techniques together with available light curves in the forthcoming study.

\section{Acknowledgements}
This study is supported by The Scientific \& Technological Research Council of Turkey (TUBITAK) with the project code 2214 and partially
by the projects TBAG106T688 and BAP2007/83 . We thank to Prof. John Hearnshaw for granting observing time and to Prof. Edwin Budding for
partial help during the project.

\begin {thebibliography}{}

\harvarditem{Andersen et al.}{1984}{A84} Andersen, J., Jensen, K. S., Nordstr\"{o}m, B., 1984, IBVS, 2642, 1
\harvarditem{Baglow}{1948}{B48} Baglow, R. L., 1948, MNRAS, 108, 343
\harvarditem{Bakis}{2007}{B07} Bak{\i}\c{s}, V., Bak{\i}\c{s}, H., Eker, Z., Demircan, O., 2007, MNRAS, 382, 609
\harvarditem{Bakis}{2008}{B08a} Bak{\i}\c{s}, V., Bak{\i}\c{s}, H., Demircan, O., Eker, Z., 2008a, MNRAS, 384, 1657
\harvarditem{Bakis}{2008}{B08b} Bak{\i}\c{s}, V., Bak{\i}\c{s}, H., Demircan, O., Eker, Z., 2008b, MNRAS, 385, 381
\harvarditem{Cerutti}{1990}{C90} Cerruti, M. A., de Laurenti, M. A., 1990, Bol. Asoc. Argent. Astron., 36, 28
\harvarditem{Cester et al.}{1978}{C78} Cester, B., Fedel, B., Giuricin, G., Mardirossian, F., Mezzetti, M.,  1978, A\&AS, 33, 91
\harvarditem{Ferrer et al.}{1986}{F86} Ferrer, O. E., Sahade, J., 1986, PASP, 98, 1342
\harvarditem{Gonzales and Lapasset}{1986}{G86} Gonzalez, J. F., Lapasset, E., 2003, A\&A, 404, 365
\harvarditem{Gum}{1951}{G51} Gum, C. S., 1951, MNRAS, 111, 634
\harvarditem{Hearnshaw et al.}{2002}{H02} Hearnshaw, J. B., Barnes, S. I., Kershaw, G. M., Frost, N., Graham, G., Ritchie, R., Nankivell, G. R., 2002, Experimental Astronomy, 13, 59
\harvarditem{Hoffleit and Jaschek}{1986}{H86} Hoffleit, D., Jaschek, C., 1991, New Haven, Conn., Yale University Observatory, c1991, 5th rev.ed., edited by Hoffleit, D., Jaschek, C.
\harvarditem{Latham et al.}{1996}{L96} Latham, D. W., Nordstr\"{o}m, B., Andersen, J., Torres, G., Stefanik, R. P., Thaller, M., Bester, M. J., 1996, A\&A, 314, 864
\harvarditem{Kholopov et al.}{1987}{K87} Kholopov, P. N., Samus, N. N., Kazarovets, E. V., Kireeva, N. N., 1987, IBVS, 3058, 1
\harvarditem{Kreiner}{2004}{K04} Kreiner, J. M., 2004, AcA, 54, 207
\harvarditem{Kurucz}{1993}{K93} Kurucz, R. L., 1993, CD-ROM 13, 18, http://kurucz.harward.edu.
\harvarditem{Manfroid and Mathys}{1996}{M96} Manfroid, J., Mathys, G., 1984, IBVS, 2616
\harvarditem{Metcalfe et al.}{1996}{Met96} Metcalfe, T. S., Mathieu, R. D., Latham, D. W., Torres, G., 1996, ApJ, 456, 356
\harvarditem{Perryman et al.}{1997}{P97} Perryman, M. A. C., 1997, A\&A, 323, 49
\harvarditem{Pearce}{1951}{P51} Pearce, J. A., 1951, AJ, 56, 137
\harvarditem{Ralchenko}{2008}{R08} Ralchenko, Yu., Kramida, A.E., Reader, J. and NIST ASD Team, 2008, NIST Atomic Spectra Database (version 3.1.5), Available: http://physics.nist.gov/asd3
\harvarditem{Robert}{1901}{R01} Roberts, A. W., 1901, AJ, 21, 81
\harvarditem{Rossiter}{1924}{R24} Rossiter, R. A., 1924, ApJ, 60, 15
\harvarditem{Sanyal and Sinyhal}{1964}{S64} Sanyal, A., Sinvhal, S. D., 1964, The Observatory, 84, 211
\harvarditem{Shobbrook}{2004}{S04} Shobbrook, R. R., 2004, The Journal of Astronomical Data, 10, 1
\harvarditem{Southworth et al.}{2007}{S07} Southworth, J. B., Clausen, J. V.,  2007, A\&A, 461, 1077
\harvarditem{Skuljan et al.}{2004}{Sk04a} Skuljan, J., Ramm, D. J., Hearnshaw, John, B., 2004, MNRAS, 352, 975
\harvarditem{Skuljan}{2004}{Sk04b} Skuljan, J., 2004, ASPC, 310, 575
\harvarditem{van Hamme and Manfroid}{1988}{V88} van Hamme, W., Manfroid, J., 1988, A\&AS, 74, 247
\harvarditem{Zucker and Mazeh}{1994}{Z94} Zucker, S., Mazeh, T., 1994, ApJ, 420, 806

\end{thebibliography}

\pagebreak

\begin{table*}
\begin{center}
\tiny
\caption{Journal of spectroscopic observations for BG~Ind.
Signal-to-noise (S/N) ratio refers to the continuum near 5500\,\AA.
\label{table1}}
\begin{tabular}{cccrccccc}
\hline
No  &  HJD        &  \multicolumn{1}{c}{Exp.Time} & S/N  & Phase & RV$_1$ & (O-C)$_1$ & RV$_2$ & (O-C)$_2$ \\
    &  (-2400000) &  \multicolumn{1}{c}{(s)}      &      &($\phi$)& (km s$^{-1}$) & (km s$^{-1}$) & (km s$^{-1}$) & (km s$^{-1}$) \\
\hline
1   & 53968.09007 & 1837 & 95  & 0.825 & 166.7 & 0.1  & -63.8 & -2.9  \\
2   & 53968.14327 & 1800 & 100 & 0.861 & 155.4 & 3.5  & -44.7 & -0.4  \\
3   & 53970.01705 & 1586 & 100 & 0.141 & -30.0 & 2.1  & 162.2 & 0.5   \\
4   & 53970.05619 & 1734 & 95  & 0.168 & -47.7 & -4.0 & 170.5 & -5.2  \\
5   & 53970.11820 & 1609 & 85  & 0.210 & -51.8 & 4.7  & 183.7 & -5.2  \\
6   & 53970.19355 & 777  & 120 & 0.262 & -52.3 & 8.0  & 196.4 & 3.1   \\
7   & 53970.22751 & 736  & 120 & 0.285 & -51.4 & 6.8  & 189.5 & -1.3  \\
8   & 53971.10174 & 994  & 115 & 0.882 & 147.0 & 6.3  & -29.0 & 4.4   \\
9   & 53972.16620 & 726  & 110 & 0.609 & 139.2 & 5.2  & -30.9 & -6.5  \\
10  & 53972.19566 & 619  & 115 & 0.629 & 148.0 & 3.2  & -35.8 & 0.6   \\
11  & 53972.22387 & 570  & 120 & 0.649 & 158.4 & 2.9  & -56.0 & -7.6  \\
12  & 53972.24855 & 575  & 120 & 0.665 & 168.2 & 6.7  & -51.6 & 3.6   \\
13  & 53976.01653 & 602  & 80  & 0.239 & -60.5 & -0.4 & 187.7 & -5.4  \\
14  & 53976.13565 & 919  & 80  & 0.320 & -52.3 & -2.7 & 180.2 & -1.2  \\
15  & 53976.16802 & 748  & 85  & 0.343 & -39.6 & 2.4  & 170.3 & -2.3  \\
16  & 53976.20147 & 851  & 75  & 0.365 & -28.3 & 0.8  & 161.0 & -0.9  \\
17  & 53977.06172 & 879  & 110 & 0.953 & 107.7 & 12.0 &   -   & -     \\
18  & 53977.10613 & 850  & 110 & 0.983 &   -   & -    & 50.6  & 6.3   \\
19  & 53977.15005 & 1198 & 105 & 0.013 &   -   & -    & 58.0  & -10.3 \\
20  & 53981.06884 & 382  & 125 & 0.690 & 169.3 & -1.1 & -69.1 & -3.9  \\
21  & 53981.10074 & 523  & 120 & 0.712 & 176.2 & 0.6  & -73.5 & -2.7  \\
22  & 53981.12565 & 495  & 110 & 0.729 & 179.1 & 1.1  & -76.5 & -2.9  \\
23  & 53982.06402 & 533  & 110 & 0.370 & -29.1 & 0.8  & 150.9 & -8.3  \\
24  & 53982.09218 & 516  & 110 & 0.389 & -17.3 & 2.2  & 141.5 & -6.1  \\
25  & 53982.11971 & 746  & 115 & 0.408 & -7.5  & 0.5  & 136.0 & 1.3   \\
26  & 53985.09603 & 1090 & 145 & 0.441 & 25.5  & 11.2 & 114.5 & 4.8   \\
27  & 53985.12285 & 901  & 145 & 0.459 &   -   & -    &   -   & -     \\
28  & 53986.07387 & 794  & 130 & 0.109 & -21.9 & -6.7 & 143.7 & 0.9   \\
29  & 53986.08455 & 862  & 135 & 0.116 & -26.4 & -6.9 & 144.2 & -3.3  \\
30  & 53987.02502 & 1018 & 135 & 0.758 & 173.2 & -5.9 & -78.0 & -3.1  \\
31  & 53987.07149 & 861  & 135 & 0.790 & 173.7 & -2.3 & -69.2 & 2.1   \\
32  & 53988.04092 & 1084 & 120 & 0.452 &   -   & -    &   -   & -     \\
33  & 53988.08411 & 925  & 135 & 0.482 & 59.1  & 15.1 &   -   & -     \\
34  & 53988.22477 & 841  & 135 & 0.578 & 119.0 & 4.7  & -0.5  & 1.7   \\
35  & 53988.97008 & 762  & 130 & 0.087 & -3.5  & -1.7 & 130.7 & 2.9   \\
36  & 53991.04171 & 714  & 130 & 0.502 & 59.8  & -0.5 &   -   & -     \\
37  & 53991.07191 & 676  & 140 & 0.522 &   -   & -    &   -   & -     \\
38  & 53991.12000 & 589  & 135 & 0.555 &   -   & -    &   -   & -     \\
39  & 53991.87132 & 761  & 135 & 0.068 & 7.1   & -3.9 &   -   & -     \\
40  & 53991.88022 & 636  & 140 & 0.075 & 11.5  & 5.4  & 125.6 & 6.7   \\
41  & 53996.05293 & 555  & 120 & 0.925 & 120.4 & 4.7  & -6.0  & -2.2  \\
\hline \\
\end{tabular}
\end{center}
\end{table*}

\begin{table*}
\begin{center}
\tiny
\caption{Journal of spectroscopic observations for IM~Mon.
Signal-to-noise (S/N) ratio refers to the continuum near 5500\,\AA.
\label{table2}}
\begin{tabular}{cccrccccc}
\hline
No  &  HJD        &  \multicolumn{1}{c}{Exp.Time} & S/N  & Phase & RV$_1$ & (O-C)$_1$ & RV$_2$  & (O-C)$_2$ \\
    &  (-2400000) &  \multicolumn{1}{c}{(s)}      &      &($\phi$)& (km s$^{-1}$)  & (km s$^{-1}$) & (km s$^{-1}$) & (km s$^{-1}$) \\
\hline
1   & 53981.18889 & 755  & 105 & 0.866 & 71.0   & 8.8   & -203.9 & -2.7  \\
2   & 53981.21547 & 698  & 105 & 0.888 & 58.8   & 11.8  & -194.0 & -5.2  \\
3   & 53981.24016 & 716  & 108 & 0.909 & 57.5   & 21.8  & -165.7 & 7.6   \\
4   & 53982.20186 & 1202 & 100 & 0.717 & 95.7   & 4.9   & -256.0 & -4.9  \\
5   & 53982.23132 & 790  & 106 & 0.742 & 112.7  & 7.3   & -241.2 & 9.0   \\
6   & 53985.19551 & 1253 & 145 & 0.232 & -153.6 & 3.7   & 199.4  & 5.6   \\
7   & 53986.18919 & 1178 & 125 & 0.067 & -57.0  & 40.9  & -      & -     \\
8   & 53986.23276 & 877  & 125 & 0.104 & -102.4 & 19.6  & 138.3  & -30.4 \\
9   & 53986.24422 & 972  & 140 & 0.113 & -109.1 & 18.8  & 141.0  & -18.7 \\
10  & 53988.19478 & 951  & 130 & 0.752 & 110.9  & 4.5   & -244.8 & 3.2   \\
11  & 53988.23537 & 954  & 130 & 0.786 & 107.3  & 5.8   & -246.1 & -1.9  \\
12  & 53989.15731 & 1020 & 135 & 0.561 & -4.0   & -19.3 & -      & -     \\
13  & 53989.18680 & 758  & 125 & 0.585 & 27.8   & 4.1   & -189.1 & -16.4 \\
14  & 53989.21414 & 756  & 125 & 0.608 & 49.9   & 7.3   & -183.7 & -11.5 \\
15  & 53989.24042 & 899  & 120 & 0.631 & 58.8   & 1.2   & -215.1 & -15.6 \\
16  & 53990.14557 & 845  & 120 & 0.391 & -130.7 & -2.6  & 87.6   & -17.2 \\
17  & 53990.17303 & 706  & 110 & 0.414 & -113.2 & -0.9  & -      & -     \\
18  & 53990.20976 & 674  & 110 & 0.445 & -66.9  & 19.9  & -      & -     \\
19  & 53991.15777 & 1101 & 125 & 0.241 & -183.2 & -2.9  & 183.7  & -10.0 \\
20  & 53991.20197 & 848  & 120 & 0.279 & -165.7 & 11.6  & 194.9  & 9.0   \\
21  & 53992.15582 & 900  & 110 & 0.080 & -121.3 & -11.2 & 78.1   & -11.5 \\
22  & 53992.20773 & 586  & 110 & 0.123 & -142.4 & -0.4  & 140.1  & 22.6  \\
23  & 53992.23900 & 693  & 110 & 0.150 & -147.3 & 9.0   & 150.0  & 7.8   \\
\hline \\
\end{tabular}
\end{center}
\end{table*}

\begin{table*}
\begin{center}
\tiny
\caption{Journal of spectroscopic observations for RS~Sgr.
Signal-to-noise (S/N) ratio refers to the continuum near 5500\,\AA.
\label{table3}}
\begin{tabular}{cccrccccc}
\hline
No  &  HJD        &   \multicolumn{1}{c}{Exp.Time} & S/N  & Phase & RV$_1$  & (O-C)$_1$ & RV$_2$  & (O-C)$_2$ \\
    &  (-2400000) &   \multicolumn{1}{c}{(s)}      &      &($\phi$)& (km s$^{-1}$) & (km s$^{-1}$) & (km s$^{-1}$) & (km s$^{-1}$) \\
\hline
1  & 53970.85740 & 289  &  95 & 0.879 & 62.7  &  7.3  & -167.5 & 20.9  \\
2  & 53970.95560 & 405  &  96 & 0.920 & 52.0  & 15.0  &   -    &  -    \\
3  & 53971.02905 & 500  &  90 & 0.950 & 60.6  & 39.6  &   -    &  -    \\
4  & 53972.05171 & 699  &  88 & 0.373 & -66.4 &  0.4  & 201.3  & 13.8  \\
5  & 53972.09327 & 642  &  78 & 0.391 & -55.0 &  3.8  & 179.9  & 16.9  \\
6  & 53978.00060 & 543  &  78 & 0.836 & 67.4  & -3.3  & -220.4 & 14.8  \\
7  & 53978.05728 & 893  &  78 & 0.859 & 54.7  &  -8.2 & -194.3 & 16.8  \\
8  & 53981.03927 & 766  &  84 & 0.094 & -51.8 &  3.1  & 128.9  & -22.0 \\
9  & 53983.80315 & 422  &  92 & 0.238 & -84.8 &  7.8  & 263.9  & -3.0  \\
10 & 53983.80857 & 402  &  89 & 0.240 & -92.1 &  0.5  & 259.2  & -7.9  \\
11 & 53983.83731 & 485  &  86 & 0.252 & -88.1 &  4.6  & 265.3  & -1.9  \\
12 & 53983.84390 & 524  &  85 & 0.255 & -86.1 &  6.5  & 266.6  & -0.4  \\
13 & 53983.88172 & 628  &  86 & 0.271 & -90.9 &  0.8  & 261.1  & -3.0  \\
14 & 53983.92387 & 1009 &  74 & 0.288 & -89.0 &  0.8  & 251.7  & -6.6  \\
15 & 53983.93586 & 948  &  75 & 0.293 & -87.0 &  2.1  & 256.4  &  0.3  \\
16 & 53984.80900 & 1204 & 112 & 0.654 & 69.7  &  0.7  & -241.5 & -11.5 \\
17 & 53984.86101 & 843  & 120 & 0.676 & 80.4  &  5.4  & -256.3 & -7.9  \\
18 & 53984.89361 & 1076 & 113 & 0.689 & 75.6  &  -2.6 & -271.1 & -12.9 \\
19 & 53984.92840 & 689  & 105 & 0.704 & 82.2  & 1.5   & -275.2 & -9.3  \\
20 & 53984.97867 & 713  & 112 & 0.725 & 91.5  & 8.4   & -275.2 & -1.7  \\
21 & 53984.98754 & 653  & 104 & 0.728 & 86.2  & 2.9   & -272.5 &  1.5  \\
22 & 53985.02380 & 764  & 106 & 0.743 & 90.8  & 6.8   & -278.0 & -2.0  \\
23 & 53985.03372 & 813  & 102 & 0.747 & 86.7  & 2.8   & -276.7 & -0.6  \\
24 & 53985.07006 & 716  &  95 & 0.762 & 90.8  & 7.1   & -272.0 &  3.0  \\
25 & 53985.07934 & 744  &  91 & 0.766 & 79.4  & -4.0  & -272.0 & 2.2   \\
26 & 53985.93479 & 648  & 118 & 0.120 & -63.2 &  2.9  & 184.5  & -1.1  \\
27 & 53986.01039 & 665  & 109 & 0.152 & -76.7 &  0.6  & 208.4  & -11.4 \\
28 & 53986.04249 & 723  &  96 & 0.165 & -78.8 &  2.7  & 221.7  & -11.0 \\
29 & 53988.86021 & 508  & 109 & 0.331 & -74.9 &  5.6  & 240.3  &  10.3 \\
30 & 53989.01592 & 568  &  98 & 0.396 & -53.1 &  3.6  & 149.1  & -7.3  \\
31 & 53989.95633 & 516  & 101 & 0.785 & 90.1  &  8.6  & -276.0 & -7.6  \\
32 & 53989.99273 & 540  &  99 & 0.800 & 79.4  &  0.1  & -274.0 & -12.5 \\
33 & 53990.02698 & 554  &  99 & 0.814 & 78.0  &  1.7  & -266.0 & -13.5 \\
34 & 53990.96969 & 1480 &  87 & 0.205 & -98.1 &  -8.6 & 255.9  & -1.7  \\
35 & 53991.89929 & 499  & 105 & 0.590 & 37.8  &  -5.8 & -163.0 & -11.3 \\
\hline \\
\end{tabular}
\end{center}
\end{table*}

\begin{table*}
\begin{center}
\caption{Spectroscopic orbital parameters of BG~Ind, IM~Mon and
RS~Sgr. \label{table5}}
\begin{tabular}{lccc}  \hline
Parameter     & \multicolumn{1}{c}{BG~Ind} &
\multicolumn{1}{c}{IM~Mon} & \multicolumn{1}{c}{RS~Sgr}\\ \hline
$P$ (days)    & \multicolumn{1}{c}{1.464069}   & \multicolumn{1}{c}{1.1902424}  & \multicolumn{1}{c}{2.4156848}  \\
$T_{0} (HJD-2453900)$  & 66.8821 $\pm$ 0.0015   & 80.1580 $\pm$ 0.0029 & 71.1436 $\pm$ 0.0038 \\
$K_{1}$(km s$^{-1}$)   & 119.9 $\pm$ 1.0  & 138.7 $\pm$ 3.1 & 88.3 $\pm$ 1.6  \\
$K_{2}$(km s$^{-1}$)   & 134.1 $\pm$ 1.0  & 228.8 $\pm$ 3.1 & 271.7 $\pm$ 1.6  \\
$e$                    & 0.0              & 0.0             & 0.0  \\
$V_\gamma$(km s$^{-1}$)&  59.4 $\pm$ 0.5  &  21.2 $\pm$ 1.8 &   -4.4 $\pm$ 1.0  \\
$q$                    & 0.894 $\pm$ 0.012& 0.606 $\pm$ 0.023 & 0.325  $\pm$ 0.004 \\
m$_{1}$sin$^{3}i (M_{\odot})$       & 1.31 $\pm$ 0.01  & 3.81 $\pm$ 0.05 & 8.81 $\pm$ 0.05 \\
m$_{2}$sin$^{3}i (M_{\odot})$       & 1.17 $\pm$ 0.01  & 2.31 $\pm$ 0.05 & 2.87 $\pm$ 0.05 \\
a$_{1}$sin$^{3}i (R_{\odot})$       & 3.47 $\pm$ 0.03  & 3.26 $\pm$ 0.07 & 4.23 $\pm$ 0.07 \\
a$_{2}$sin$^{3}i (R_{\odot})$       & 3.88 $\pm$ 0.03  & 5.38 $\pm$ 0.07 & 12.98$\pm$ 0.07 \\
r.m.s. (km s$^{-1}$)              &  4.4   &  10.7 & 7.9   \\
\hline
\end{tabular}
\end{center}
\end{table*}

\begin{figure}
\begin{center}
\begin{tabular}{c}
      \resizebox{100mm}{100mm}{\includegraphics{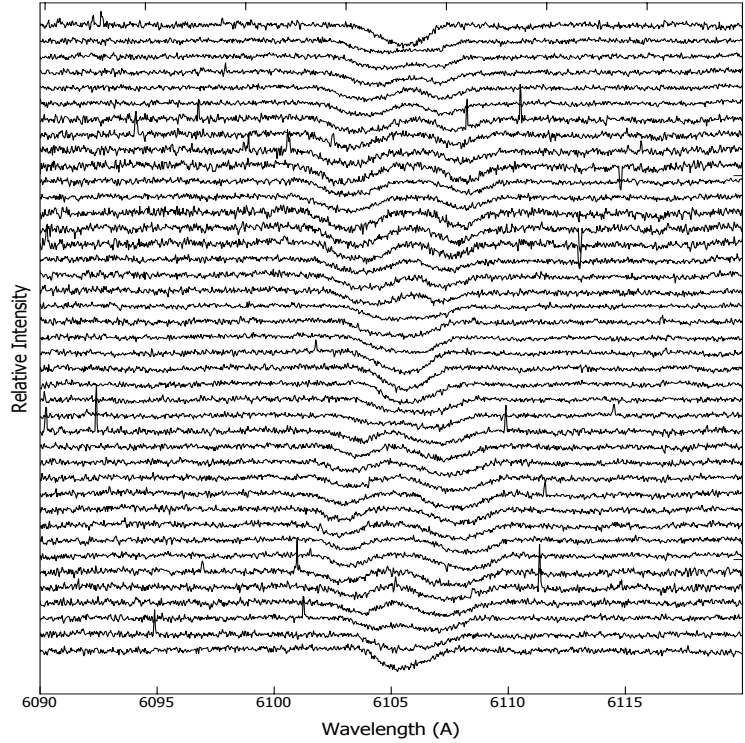}} \\
\end{tabular}
\caption{The Ca I (6102.723 \AA) lines of the components of BG~Ind
used for RV measurements. The spectra are ordered } \label{fig1}
\end{center}
\end{figure}

\begin{figure}
\begin{center}
\begin{tabular}{c}
      \resizebox{100mm}{100mm}{\includegraphics{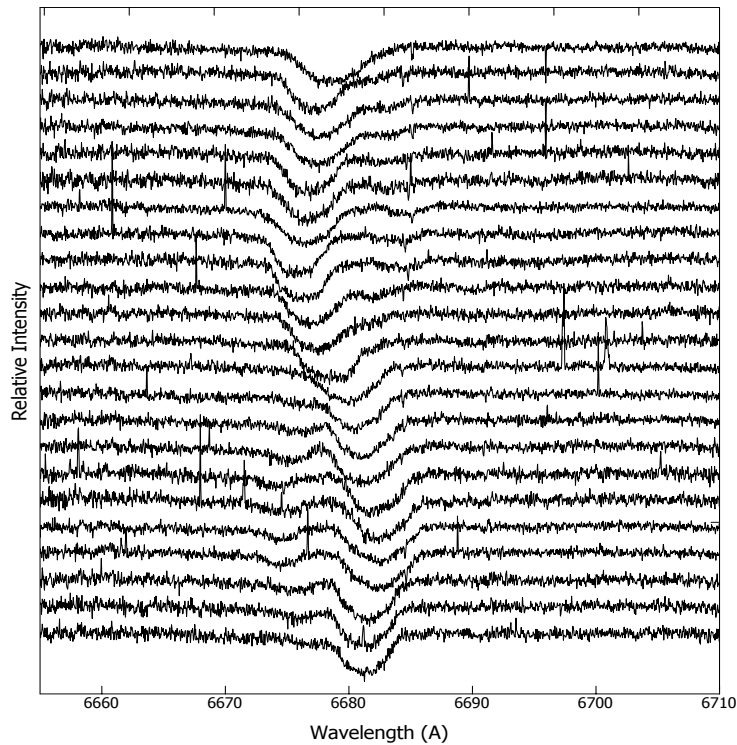}} \\
\end{tabular}
\caption{The He I (6678 \AA)\,lines of the components of IM~Mon used
for RV measurements.} \label{fig2}
\end{center}
\end{figure}

\begin{figure}
\begin{center}
\begin{tabular}{c}
      \resizebox{100mm}{100mm}{\includegraphics{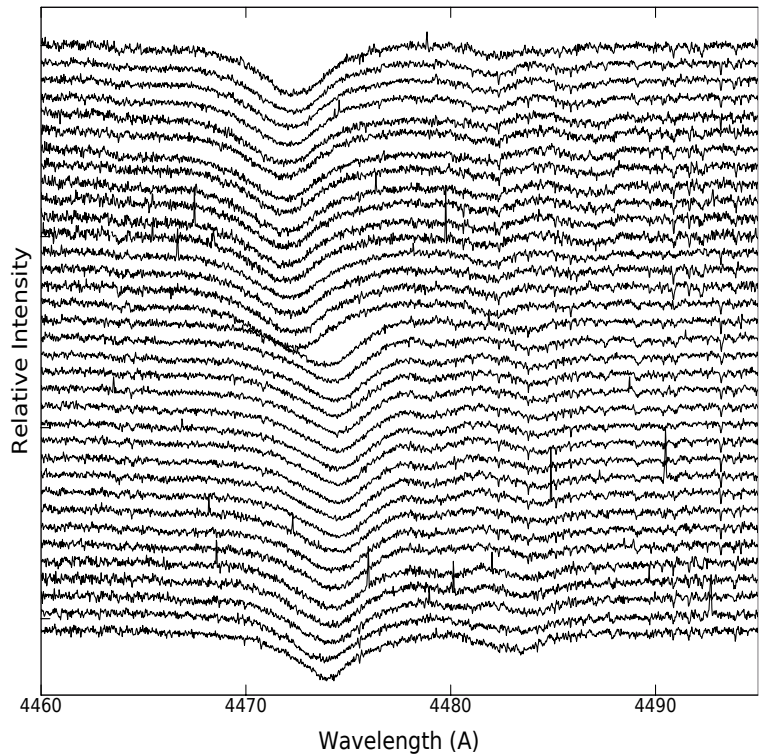}} \\
\end{tabular}
\caption{The He I (4471 \AA)\,and Mg II (4481 \AA) lines of the
components of RS~Sgr used for RV measurements.} \label{fig3}
\end{center}
\end{figure}

\begin{figure*}
\begin{center}
\begin{tabular}{c}
      \resizebox{100mm}{!}{\includegraphics{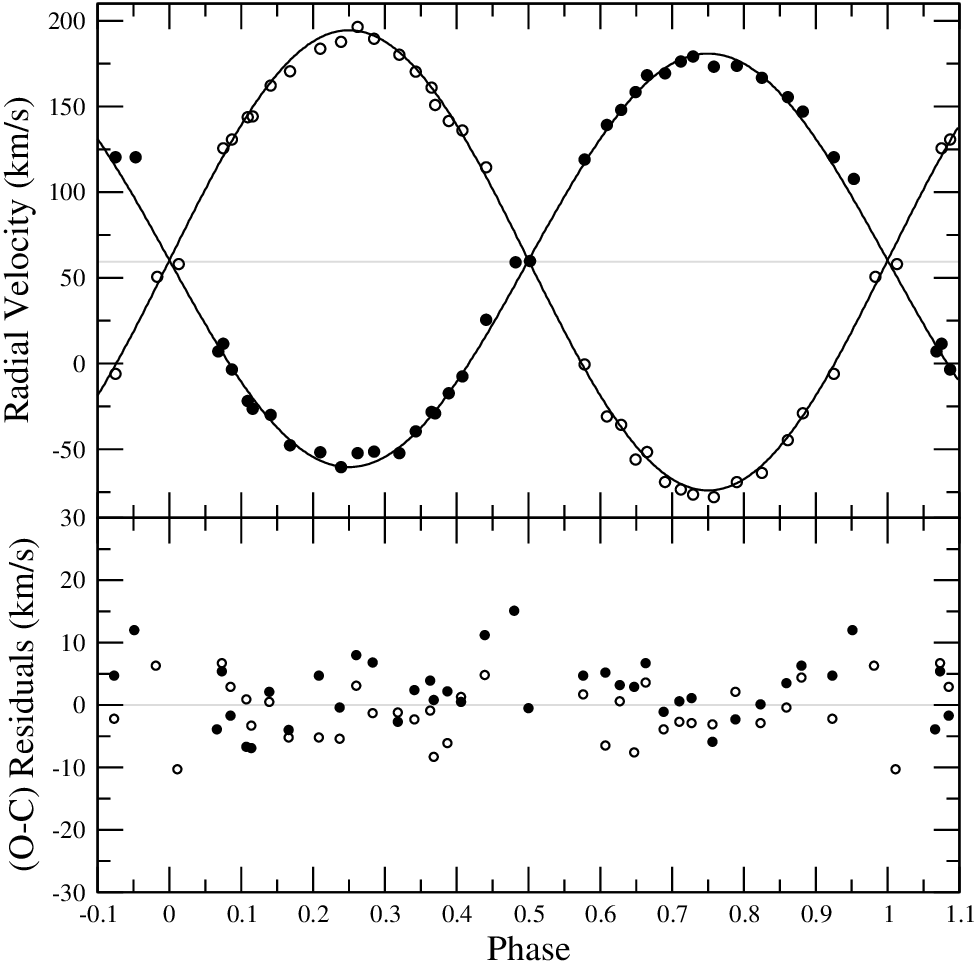}} \\
\end{tabular}
\caption{Spectroscopic orbital solution for BG~Ind and radial
velocities (filled circles: primary, open circler: secondary). The
dotted line in the figure represents the center of mass velocity.}
\label{fig4}
\end{center}
\end{figure*}

\begin{figure*}
\begin{center}
\begin{tabular}{c}
      \resizebox{100mm}{!}{\includegraphics{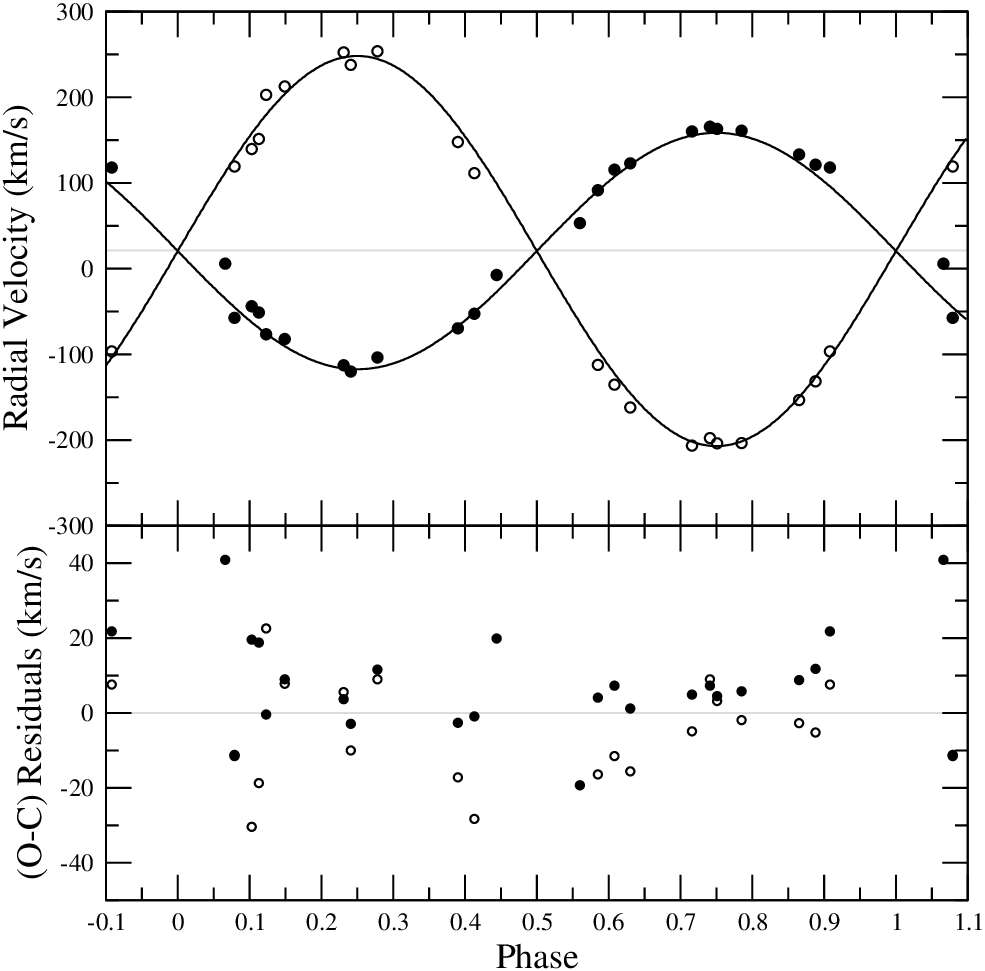}} \\
\end{tabular}
\caption{Spectroscopic orbital solution for IM~Mon and radial
velocities (filled circles: primary, open circler: secondary). The
dotted line in the figure represents the center of mass velocity.}
\label{fig5}
\end{center}
\end{figure*}

\begin{figure*}
\begin{center}
\begin{tabular}{c}
      \resizebox{100mm}{!}{\includegraphics{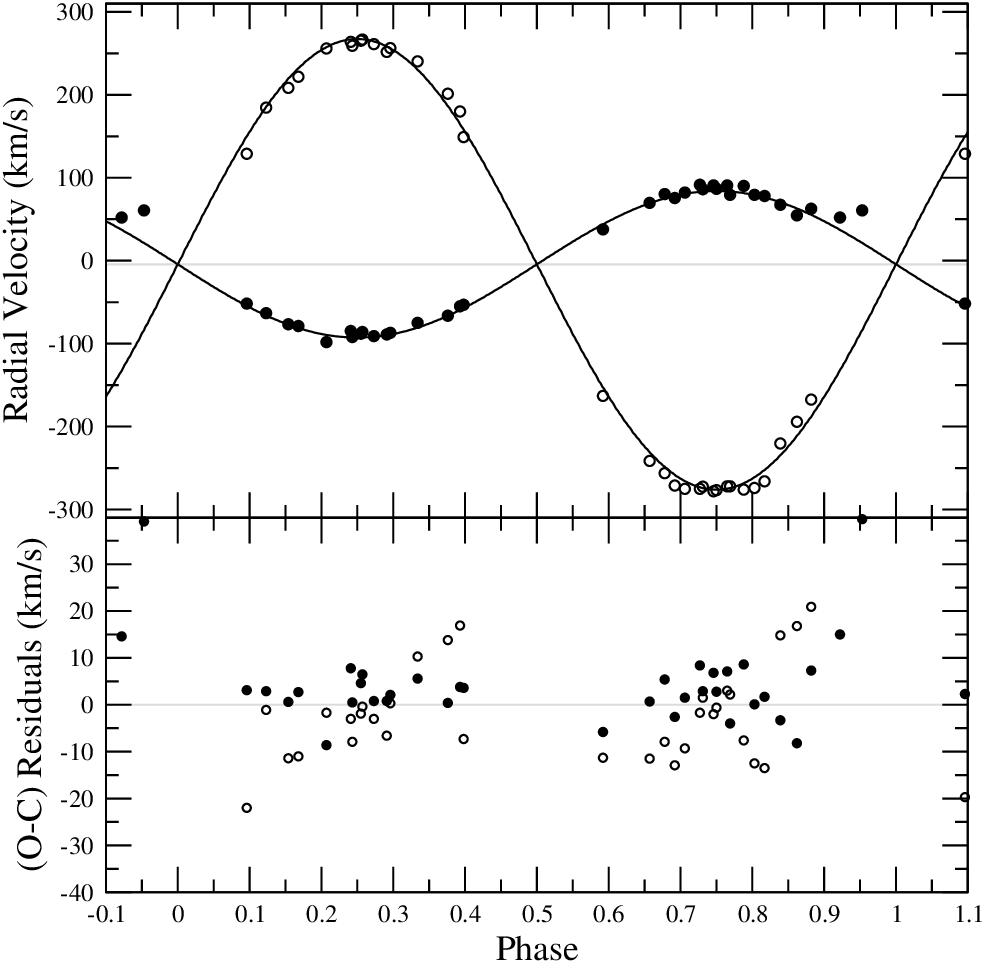}} \\
\end{tabular}
\caption{Spectroscopic orbital solution for RS~Sgr and radial
velocities (filled circles: primary, open circler: secondary). The
dotted line in the figure represents the center of mass velocity.}
\label{fig6}
\end{center}
\end{figure*}

\end{document}